\documentclass[runningheads,a4paper]{llncs}

\usepackage{amssymb}
\usepackage{graphicx}
\usepackage{pseudocode}
\usepackage{amstext}
\usepackage{url}
\usepackage{subfigure}

\urldef{\mailsa}\path|{alfred.hofmann, ursula.barth, ingrid.haas, frank.holzwarth,| \urldef{\mailsb}\path|anna.kramer, leonie.kunz, christine.reiss,
nicole.sator,| \urldef{\mailsc}\path|erika.siebert-cole, peter.strasser, lncs}@springer.com|
\newcommand{\keywords}[1]{\par\addvspace\baselineskip
\noindent\keywordname\enspace\ignorespaces#1}

\begin{document}

\mainmatter  

\title{Knowledge Network System (KNS) by Evolutionary Collective Intelligence (ECI): Model, Algorithm and Applications}

\titlerunning{KNS by ECI}

%
%
\author{Tao Xiang, Ziliang Huang, Peng Bai, Congrui Ji, and Zhiyong Liu\footnote{Corresponding author}}
\authorrunning{Tao Xiang, Ziliang Huang, Peng Bai, Congrui Ji, and Zhiyong Liu}


\institute{Visva, Inc.
20410 Town Center Lane, Suite 295
Cupertino, CA 95014, U.S.A.\\
\email{\{taoxiang, zilianghuang, pengbai, congruiji, zhiyongliu@visva.com \}}}

%
%

\toctitle{Lecture Notes in Computer Science} \tocauthor{Authors' Instructions}
\maketitle

\begin{abstract}
Aiming at overcoming some inherent drawbacks and bottlenecks encountered by the conventional Knowledge, Recommendation, Search, and Social Systems, in this article we introduce the Knowledge Network System (KNS), a novel type of knowledge graph which is constructed by a new proposed algorithm, the Evolutionary Collective Intelligence (ECI) algorithm. The ECI, an agent-machine interactive algorithm, constructs the KNS by iteratively recommending interesting/matched samples/files to the agents, and meanwhile taking advantages of the collective intelligence of the agents. The ECI based KNS, to the best of our knowledge, is the first attempt in literature that integrates the functions of knowledge network construction, high-quality recommendation, new types of search and social in a same framework. Some real and potential applications of KNS and ECI are discussed, and a real system named VISVA is provided to demonstrate their efficacy. Some open problems for future works are also summarized in the end.

\end{abstract}

\section{Introduction}
Nowadays, people query/search/receive information from the Internet by using a variety of tools, Knowledge System such as Wikipedia, Search Engine such as Google, Recommend Systems such as TopBUZZ, and Social Network System (SNS) such as Twitter, to name a few. The underlying technologies behind them have proven quite effective during the last decade. Generally, the Knowledge System employs entry to facilitate the query, Search Engines make use of key words matching, Recommend System works by mining the correlations between the articles and/or users, and SNS relies on the follow relationship between users. Though these tools have achieved great success by now, all of them exhibit some inherent drawbacks.

The entry system makes query convenient to implement, but it somehow suffers from at least the two facts. First, the entry system is tedious to maintain, especially in nowadays many new ideas and concepts emerge almost everyday; second, it is hard to precisely locate in the entry system some multidisciplinary knowledge, let alone those concepts beyond words and unnameable. For the key words based search engines, only the articles containing the key words can be found, but cannot hit those closely related materials but without explicitly containing the key words; it is even harder, if not impossible, to recover those nonverbal medias such as image, video and audio files. Huge number of recommendation algorithms have been proposed in recent years. The content or user based recommendation recommends the content based on some \emph{similarity} between some related contents or users, but hard to generalize to find other potential interesting items; the knowledge graph based recommendation can work in a more systematic way, but how to formulate the knowledge graph (especially suitable for recommendation) remains to be a tough task. The follow relationship in SNS means that the follower receives all of the information posted by the one followed, however, in general not all but only a part of the posts might be interesting to the follower, consequently making the follower receive much \emph{noise}.

In view of the above considerations, in this paper we introduce the Knowledge Network System (KNS), which is gradually formulated by the Evolutionary Collective Intelligence (ECI), a novel agent(human)-machine interactive learning algorithm proposed particularly for the KNS. The main contribution of this article is summarized as twofold:
\begin{itemize}
  \item To overcome the drawbacks of the knowledge and recommendation systems etc., we propose the ECI algorithm to construct the KNS, a new type of knowledge construction, storage and distribution system.
  \item A real application based on the KNS and ECI is presented to illustrate their efficacy.
\end{itemize}

The remainder of this article is organized as follows. The KNS and ECI algorithms are introduced in Sections \ref{sec_2} and \ref{sec_3}, respectively, followed by some theoretical analysis and discussions in Section \ref{sec_4}. After giving some real and potential applications in Section \ref{sec_5}, Section \ref{sec_6} concludes this article by summarizing some future works.

\section{Knowledge Network Systems}
\label{sec_2}
Some notations used in the KNS and ECI algorithm are firstly given as follows.
\begin{itemize}
  \item \emph{Knowledge file}, denoted by $f$, is the basic input or sample of the knowledge system. Typical types of knowledge file include such as message, article, image, video, and audio file.
  \item \emph{Knowledge item}, denoted by $v$,  exactly a knowledge \emph{node} in the KNS, represents one basic knowledge point in the KNS. Each item consists of one or more knowledge files, or in other word, the knowledge item is a collection of these files, which share some common characteristics. Since as will see below each node in the KNS is also accompanied with some agents, sometimes we may also call each \emph{item} a \emph{group} in the sense of group of agents. That is, \emph{item}, \emph{node}, and \emph{group} can be exchanged equivalently by context.
  \item \emph{Agent}, denoted by $a$, might being a person or an artificial agent, is defined as an intelligent reactor that might be \emph{interested} to some particular files. If an agent $a$ is interested in a file $f$ pushed to it, we say that $f$ \emph{matches} $a$, and $f$ is a \emph{signal} to $a$, or a \emph{noise} otherwise. Thus, for each agent $a$,  we can define a signal-noise ratio (SNR) on the pushed files as follows,
      \begin{equation}\label{eqn:snr_a}
      snr_a= \frac{\#\text{matched files}}{\#\text{unmatched files}}
      \end{equation}

\end{itemize}

\begin{figure}[htp]
\centering
   \scalebox{1}[1]{\includegraphics[scale=0.5]{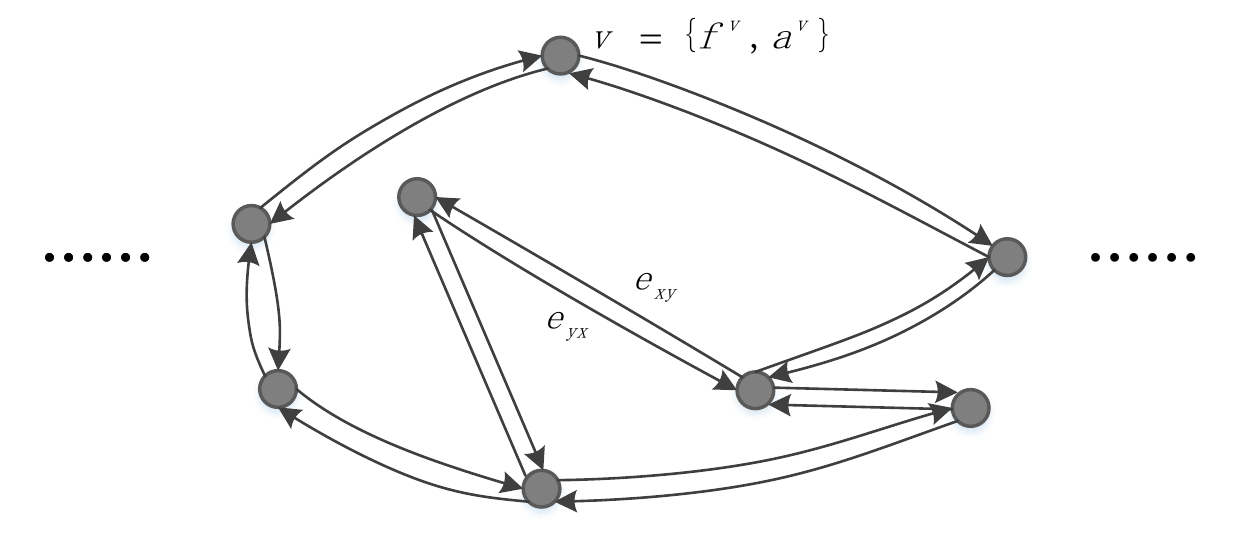}}
\caption{A simple illustration of KNS}\label{Fig:kns}
\end{figure}

The Knowledge Networks Systems (KNS), denoted by $\mathcal{K}$, is defined as a dynamic directed knowledge graph $\mathcal{K}=\{V, E\}$, where $V$ is the set of graph nodes and $E=(v_x,v_y)$ defines the directed edges from node $v_x$ to $v_y$, as illustrated by Fig. \ref{Fig:kns}. Here \emph{dynamic} implies that the graph including both $V$ and $E$ may evolve over time. Actually, as will see below, $\mathcal{K}$ could be evolved initially from a null graph without containing any node. Without doubt, KNS is also possible to be a static graph with fixed nodes and structure if the evolution is finished.

Each node $v$ in the KNS represents a knowledge item, and $v$ has two attributes. The first one is the knowledge files it consists of, and the second one is the agents belonging to it. So, $v$ can be represented as
\begin{equation}\label{eqn_v}
  v=\{f^v,a^v\}, f^v=\{f_1^v,f_2^v,...f_k^v\}, a^v=\{a^v_1,a^v_2,...,a^v_n\}
\end{equation}
It is noted that in general it is $f^v$ instead of $a^v$ that defines the knowledge item, just because each agent may cover huge number of knowledge items. It is also because as shown in the next section the collected files $f^v$ in each $v$ are voted together by the agents, the knowledge item takes a quite different form from the traditional entry.

Based on $f^v$ and given two nodes $v_x$ and $v_y$, the weight value of the edges $e_{xy}$ and $e_{yx}$ can be calculated respectively as
\begin{equation}\label{ean_exy}
e_{xy}=\frac{|f^{vx}\cap f^{vy}|}{|f^{vy}|}, e_{yx}=\frac{|f^{vx}\cap f^{vy}|}{|f^{vx}|}.
\end{equation}

Each directed weighted edge between two nodes measures not only the similarity, but also a hierarchical relationship between them. For example, given two directed edges $e_{xy}=0.9$ and $e_{yx}=0.5$ , the similarity between the two nodes can be directed calculated by $\frac{1}{2}(e_{xy}+e_{yx})=0.7$, and $v_x$ positions a higher hierarchical level in $\mathcal{K}$ since simply $e_{xy} > e_{yx}$.
In addition, once the KNS is established, we may also employ some hierarchical spectral clustering tools \cite{von2007tutorial,sanchez2014hierarchical} on the KNS to mine its hierarchical knowledge structure.

In practice, we may set a threshold value $T_e$ for the weighted edges, that is, $e_{xy}$ is set to be zero if $e_{xy} < T_e$. Thus, the KNS is in general a sparse graph, meaning that only a very small fraction of the node pairs is activated. This sparse property is particulary important for a large $\mathcal{K}$, considering both the storage and computational complexities.

\section{Evolutionary Collective Intelligence} \label{sec_3}
In this section we introduce the Evolutionary Collective Intelligence (ECI), an agent-machine interactive algorithm proposed particularly to generate the KNS.

\begin{figure}[htp]
\centering
   \scalebox{1}[1]{\includegraphics[scale=0.5]{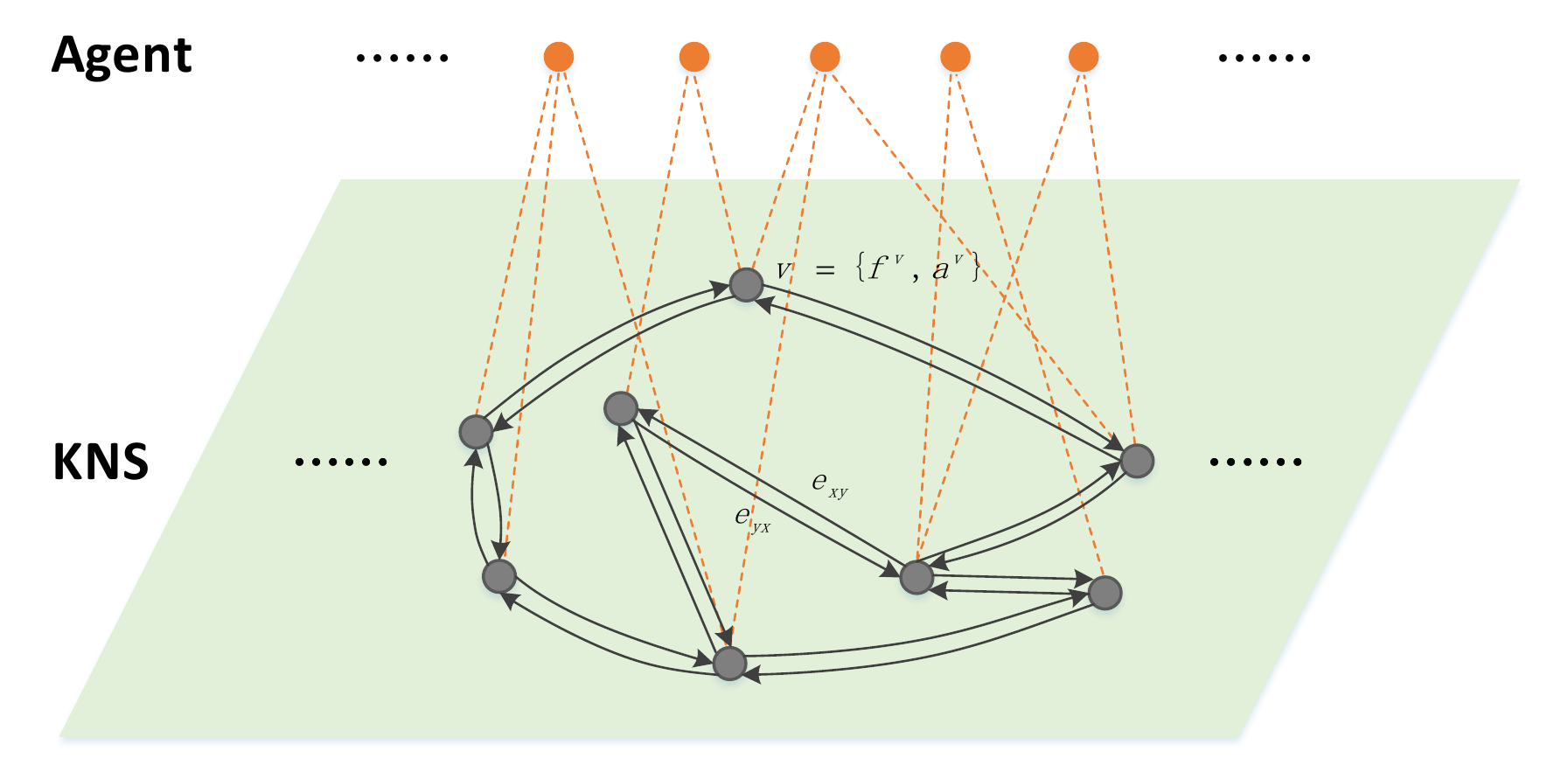}}
\caption{A simple illustration of the interaction between agents and KNS}\label{Fig:kns_agent}
\end{figure}

Generally, the ECI algorithm works by interactions between the agents and the machine: the machine recommends the files to the agents, and the agent chooses to activate or not on the recommendation. The relationship between the agents and the KNS is illustrated by Fig. \ref{Fig:kns_agent}, where the top and bottom layers are the agents and KNS, respectively. If an agent links to some nodes of the KNS, it means that the agent matches this items or belongs to this group. And the ECI algorithm is implemented by extensively using such relationships between the agents and KNS. Actually, the goal of the ECI algorithm comprises two parts: first to generate the KNS with each item consisting of multiple files sharing comment knowledge, and second to achieve a high-qualify recommendation, i.e., high $snr_a$ for agents.

\begin{figure}[htp]
\centering
   \includegraphics[scale=0.3]{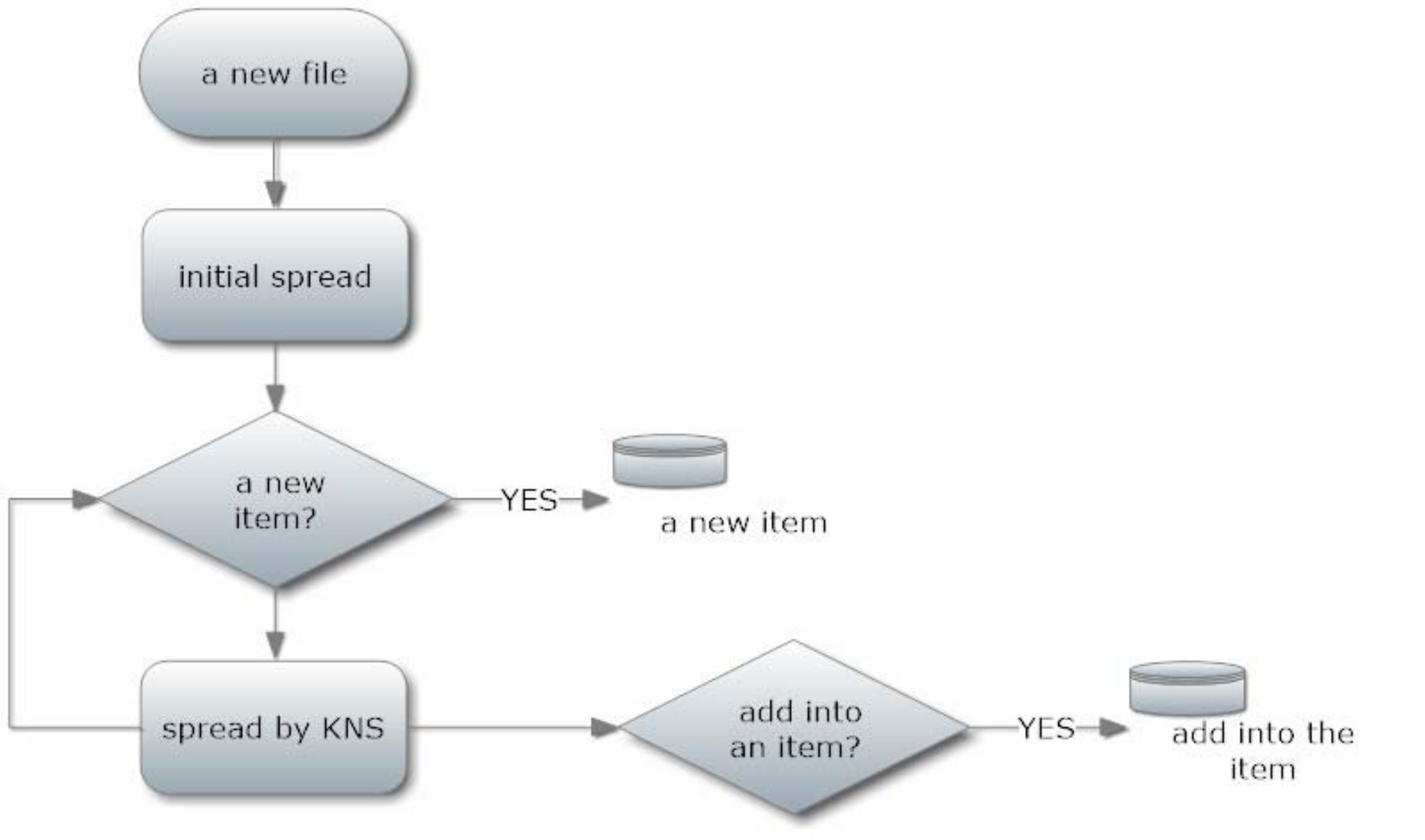}
\caption{The basic flowchart of ECI algorithm}\label{Fig:algchart}
\end{figure}

The basic flowchart of ECI algorithm is shown in Fig. \ref{Fig:algchart}, where the algorithm starts by a new generated file $f$, posted by the system or an agent. Then $f$ is recommended/pushed to some initially selected agents $\hat{A}$ (\emph{initial spread} in the flowchart), usually in a random or other sophisticated way. More importantly, except for the initial spread, $f$ will also be pushed to some or all of the agents belonging to those activated $items$ $\hat{I}$ (\emph{spread by KNS} in the flowchart). Meanwhile, $f$ is checked to be qualified or not to be added into $\hat{I}$ (\emph{add into the item} in the flowchart), and whether qualified or not, as the founding $file$, to generate a new item (\emph{new item} in the flowchart).

\emph{Spread by KNS} is the core module of the algorithm which are crucial to achieve the two goals mentioned above, i.e., to formulate the KNS and to get a high SNR. We give some more details about the module below.

In ECI algorithm, a file $f$ is pushed to the agents by two ways: \emph{initial spread} and \emph{spread by KNS}. Because the \emph{initial spread} is usually implemented by randomly selecting some initial agents for a new file, its SNR is basically around the random recommendation SNR. Therefore, the SNR is improved based mainly on the \emph{spread by KNS}, for which two mechanisms are proposed to achieve a high quality recommendation: voting mechanism and multi-round spread mechanism. The voting mechanism refers to that to make a file qualified or further spread in an item, most agents in a group should \emph{match} the file, or in other words, most agents vote for the file (voting ratio should be bigger than a preset threshold). The multi-round spread mechanism means that a file spreads in a group by being pushed to only some but not all agents in the group in each round, and the file will not spread further in the next round unless it successfully passes the agents voting in the current round. The former one guarantees that only those matched files to the agents could be qualified or even spread in their group, and the latter one further ensures a high SNR by a gradually progressive sampling mechanism.

Some other modules are briefly summarized here. \emph{New item} means that if the number of matched agents of a file $f$ is bigger than a threshold, a new item $v$ is created by taking $f$ as the founding file of $v$, and also some matched agents are added into the item, i.e., $v_a$ in eqn.(\ref{eqn_v}). \emph{Add into the item} is to add a file into an item, i.e., $v_f$ in eqn.(\ref{eqn_v}). Normally, the file that passes the whole multi-round spread in an item/group and meanwhile satisfies some other measures\footnote{For example, the degree of matching between $f$ and the files already in an item $v$ should be checked before $f$ can be added into $v$.} will be added into the item.

\section{Theoretical Analysis and Discussions} \label{sec_4}
An overview of the KNS and ECI was described above. Next we give some theoretical analysis and discussions on them.
\subsection{on agent}
In the ECI algorithm, a key step is to check if an agent $a$ matches a file $f$ or not. In the case of the agent being a human user, when the file is pushed to a user, the \emph{match} is embodied by some positive actions, such as \emph{click}, \emph{comment}, or \emph{save} of the user on the file. But if the agent is an artificial but non-human agent, we need to introduce extra information, on both agent and file, to make the algorithm proceed. Below an extra variable \emph{knowledge unit} $u$ is introduced for the artificial agent case. Though the concept of \emph{knowledge unit} is unnecessary for a human user system to implement, it, as an implicit hidden variable, still helps to theoretically analyze both KNS and ECI.

A \emph{knowledge unit}, denoted by $u$, is considered as the most basic knowledge element in the knowledge system, which cannot be further decomposed. In this sense, a knowledge \emph{file} $f$ can be taken as a collection of one or more \emph{units}; meanwhile, from the viewpoint of \emph{agent}, the \emph{knowledge unit} can be regarded as the \emph{interest} of the agent, and thus an agent $a$ can also be taken as a collection of $interests$ or $units$, That is, given a knowledge unit set
\begin{equation}\label{eqn_u}
  U = \{u_1,u_2,....,u_m\},
\end{equation}
a knowledge file $f$ and an agent $a$ can be respectively expressed as a $m$-dimensional $0-1$ vectors
\begin{eqnarray}\label{eqn_fa}
f&=&[\tau^f_1,\tau^f_2,....,\tau^f_m]^T \nonumber \\
a&=&[\tau^a_1, \tau^a_2,..., \tau^a_m]^T
\end{eqnarray}
where $\tau=\{0,1\}$, and $\tau_i=1$ implies $f$ or $a$ consists of the $i$th \emph{knowledge} or \emph{interest} unit, i.e., $u_i$.

Then, based on eqn. (\ref{eqn_fa}), the \emph{match} between $a$ and $f$ is directly checked by the following $r(a,f)$,
\begin{equation}\label{eqn_raf}
   r(a,f)=\left\{
   \begin{array}{rcl}
       0 & &\text{if }  f^Ta = 0,\\
       1 &  &\text{otherwise}
    \end{array}
           \right.
\end{equation}
Eqn. (\ref{eqn_raf}) implies that if $f$ and $a$ share at least one unit, then $a$ matches $f$. We can also assign the number of units shared to $r(a,f)$ to measure the degree of match. In addition, $r(a,f)$ can be regarded as a SNR, meaning that the file $f$ is a signal to the agent $a$ if $r(a,f)=1$, or noise otherwise. It is noted that such a definition is consistent with the SNR defined in eqn. (\ref{eqn:snr_a}). In practice, $a$ or $f$ defined in eqn. (\ref{eqn_fa}) are usually sparse vectors, especially when $m$ is large.

\subsection{on objective of ECI}
As mentioned above, the goals of ECI are twofold: the KNS and a high SNR, meaning that the KNS can be gradually evolved along with a high-qualify recommendation system. So the two types of systems, knowledge and recommendation systems, mutually promote each other.

For the KNS, each of its node represents one knowledge item, via a collection of knowledge files. So, these files should be cohesive to make the item compact and focused. The cohesion of each item can be measured by the mean square reconstruction error (MSRE) \cite{xu1993least} of the files: the smaller the error the more cohesive the item. Given a knowledge item $v$ and its $f^v=\{f_1^v,f_2^v,...f_k^v\}$ defined by eqn. (\ref{eqn_v}), the MSRE of $v$ is given by
\begin{equation}\label{eqn_msre_node}
msre(v) = \frac{1}{k}\sum_{i=1}^{k}(f_i^v - \overline{f^v})^T(f_i^v - \overline{f^v})
\end{equation}
where $\overline{f^v} = \frac{1}{k}\sum_{i=1}^{k}f_i^v$ is the mean of $f^v$. Based on the MSRE of each item, and by denoting the item set $V=\{v_1,v_2,...,v_l\}$ of the KNS $\mathcal{K}$, the MSRE or the objective function for $\mathcal{K}$ is formulated as follows,
\begin{equation}\label{eqn_msre_kns}
  MSRE(\mathcal{K})= \frac{1}{l} \sum_{i=1}^{l} msre(v_i).
\end{equation}

Minimization of $MSRE(\mathcal{K})$ defines a typical clustering problem, i.e., MSRE clustering which is inherently a NP-hard problem. So in this sense, the ECI provides an algorithm to approximately solve this NP-hard problem.

It is noted that, in literature, K-means algorithm \cite{hartigan1979algorithm} is a commonly used technique for solving such clustering problem. However, the K-means like algorithm is inapplicable here, because in practice it is usually hard to explicitly get the \emph{knowledge unit} set $U$, which is introduced here as a hidden variable for theoretical analysis.

For the second goal, a high SNR, the SNR of a file $f$ to an agent $a$ is given by eqn.(\ref{eqn_raf}). Thus, by denoting $f^a=\{f^a_1, f^a_2, ..., f^a_t\}$ as the file list pushed to $a$, the SNR for $a$ is given as
\begin{equation}\label{eqn_snra}
  snr(a) = \frac{1}{t}\sum_{i=1}^{t} r(a,f^a_i)
\end{equation}
And for the agent set $A=\{a_1,a_2,....,a_n\}$, the system SNR can be averagely gotten by
\begin{equation}\label{eqn_snrA}
  SNR(A) = \frac{1}{n}\sum_{i=1}^{n} snr(a_i)
\end{equation}

Generally, it is difficult to simultaneously minimize $MSRE(\mathcal{K})$ and maximize $SNR(A)$. ECI, to our best knowledge, is the first attempt in literature to target at the two goals simultaneously.

\subsection{on comparison with related techniques}
The ECI realizes recommendation and constructs knowledge system, but in a much different way from conventional techniques. ECI based KNS can avoid some troubles faced by nowadays related systems, including not only \emph{recommendation} and \emph{knowledge}, but also \emph{search} and \emph{social} systems. Below we give a brief comparison between KNS and the conventional systems. it is noted that for the above four types of systems, the agents in the ECI and KNS are all human user.

From the viewpoint of \emph{knowledge}, by the ECI, the KNS does not need to maintain a sophisticated but tedious entry system; also it can naturally adapt to knowledge changes, including new emerged concepts; furthermore, compared to the tree structure used by encyclopedia, the KNS is much more flexible on mining the knowledge structure.

From the viewpoint of \emph{recommendation}, by the KNS, the system can recommend closely related files to users based on the knowledge structure, but not just rely on the similarity (by means of content or users) based techniques. In addition, by the multi-round voting mechanism, the KNS recommendation achieves a high SNR, i.e., a high-qualify recommendation.

From the viewpoint of \emph{search}, the KNS system can provide not only articles, but also any other types of medias, including typically image, video and audio files. This is by now extremely difficult for conventional search engines. Also, somehow similar to $recommendation$, the search results can include those closely related contents without needing to explicitly contain the searching keywords.

From the viewpoint of \emph{social}, as illustrated by Fig. \ref{Fig:kns_agent}, in KNS the users are not connected directly, but indirectly via the knowledge or interest items. This structure truly realizes the \emph{interest} social and avoids the hard connection between users, commonly used by SNS systems. In such as \emph{interest} social system, one user receives only commonly interesting files posted by other users, but excludes the \emph{noises}.

\section{Applications}
\label{sec_5}
In this section we describe some real and potential applications of the ECI based KNS.
\subsection{VISVA}
The VISVA system\footnote{\url{http://www.visva.com} and \url{http://www.visvachina.com}.} is a knowledge/interest construction, share and recommendation platform, where the posted files will be pushed to interested users, and users receive their interesting recommendations. With the help of the KNS and ECI, the VISVA system integrates all of the four functions: \emph{knowledge}, \emph{recommendation}, \emph{search} and \emph{social}, in a different way from the traditional experiences.

\begin{figure}[htbp]
\centering
\subfigure{
\begin{minipage}[t]{0.25\linewidth}
\centering
\includegraphics[bb=0 0 2560 1700,width=2.2\textwidth,clip]{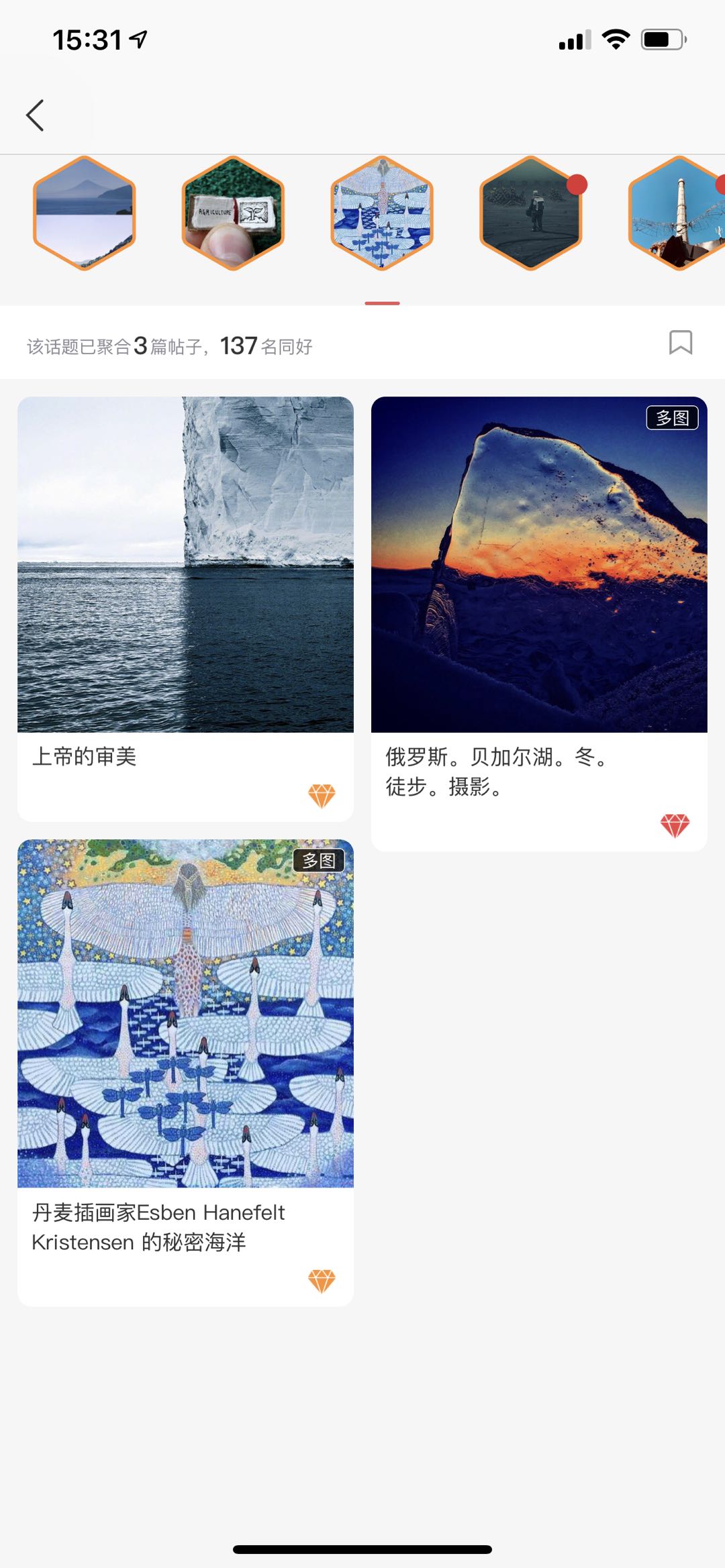}
\end{minipage}%
}%
\subfigure{
\begin{minipage}[t]{0.25\linewidth}
\centering
\includegraphics[bb=0 0 2560 1700,width=2.2\textwidth,clip]{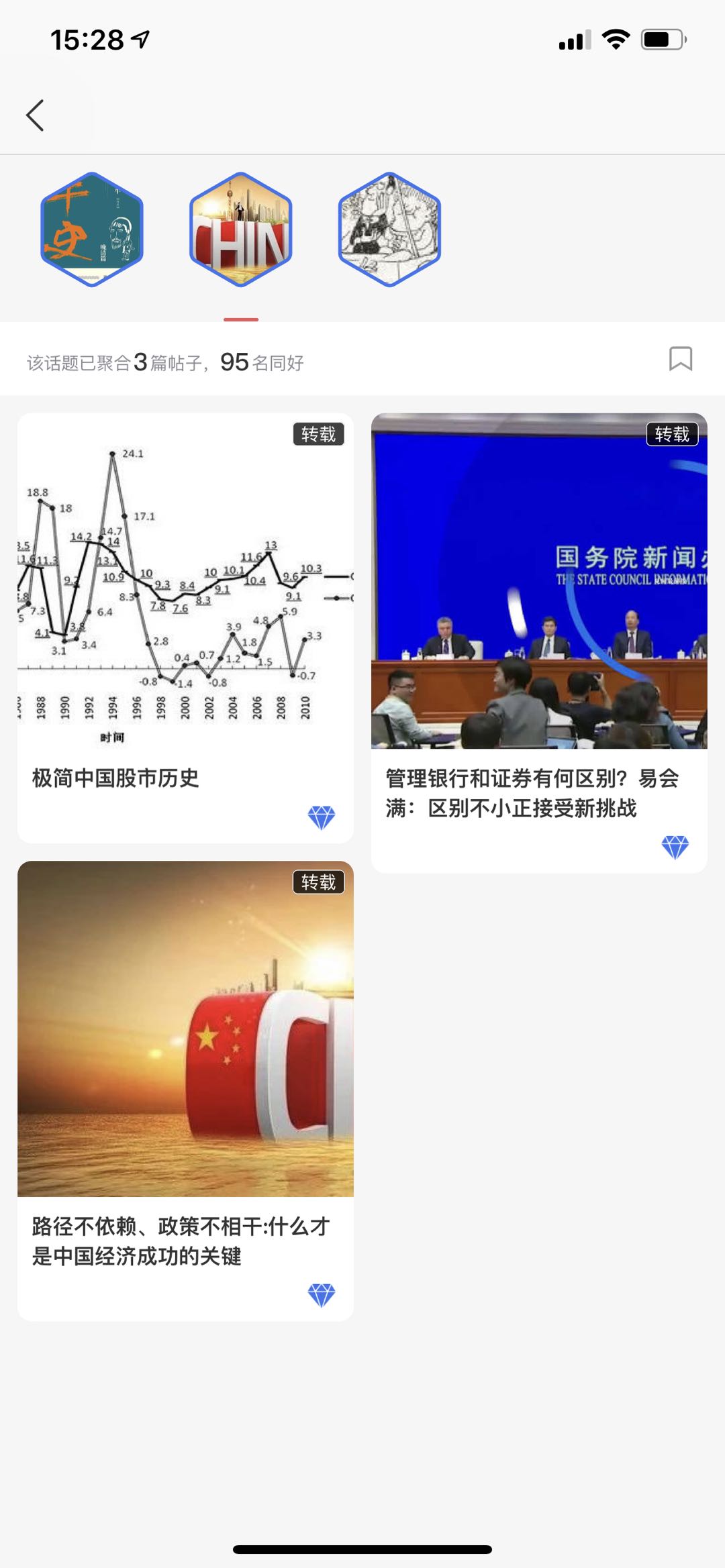}
\end{minipage}%
}%
\subfigure{
\begin{minipage}[t]{0.25\linewidth}
\centering
\includegraphics[bb=0 0 2560 1700,width=2.2\textwidth,clip]{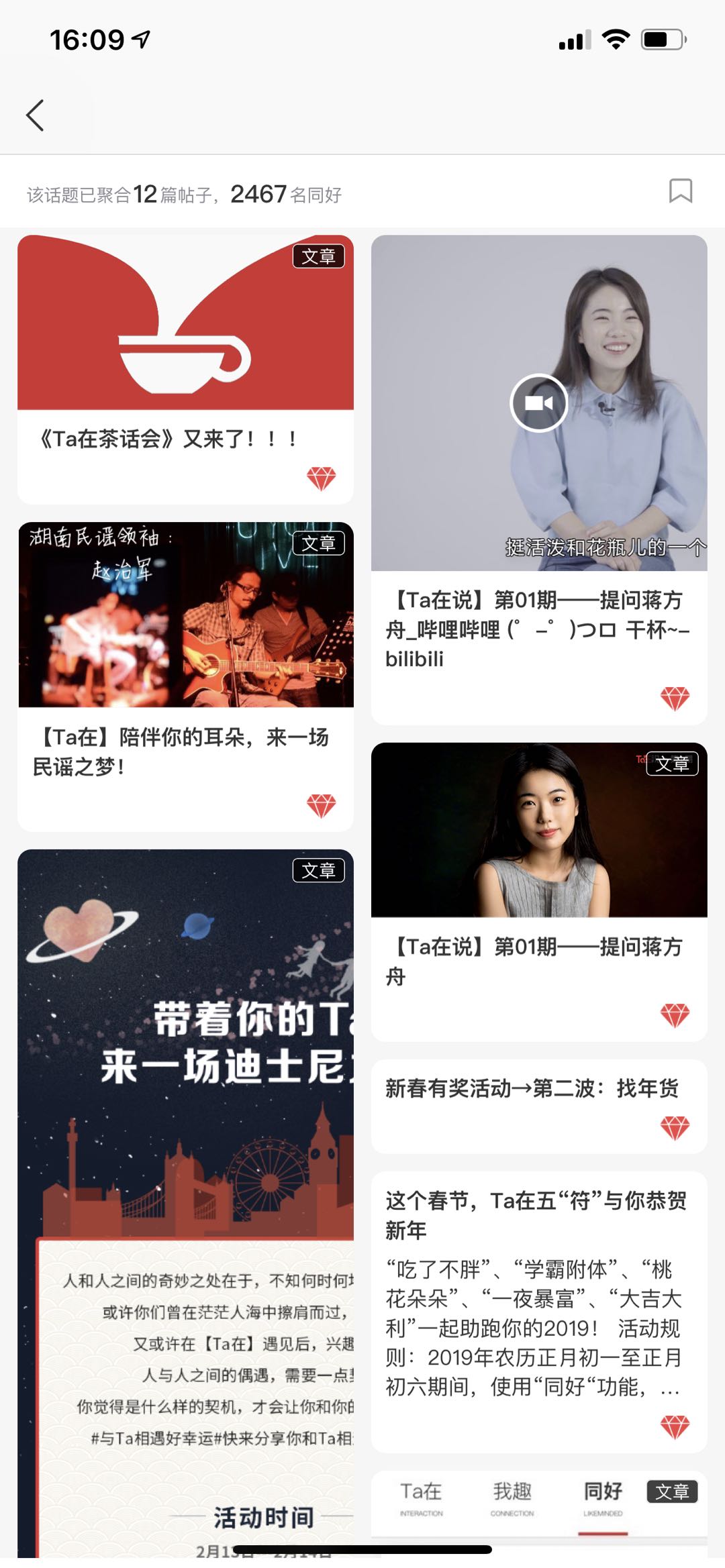}
\end{minipage}
}%
\subfigure{
\begin{minipage}[t]{0.25\linewidth}
\centering
\includegraphics[bb=0 0 2560 1700,width=2.2\textwidth,clip]{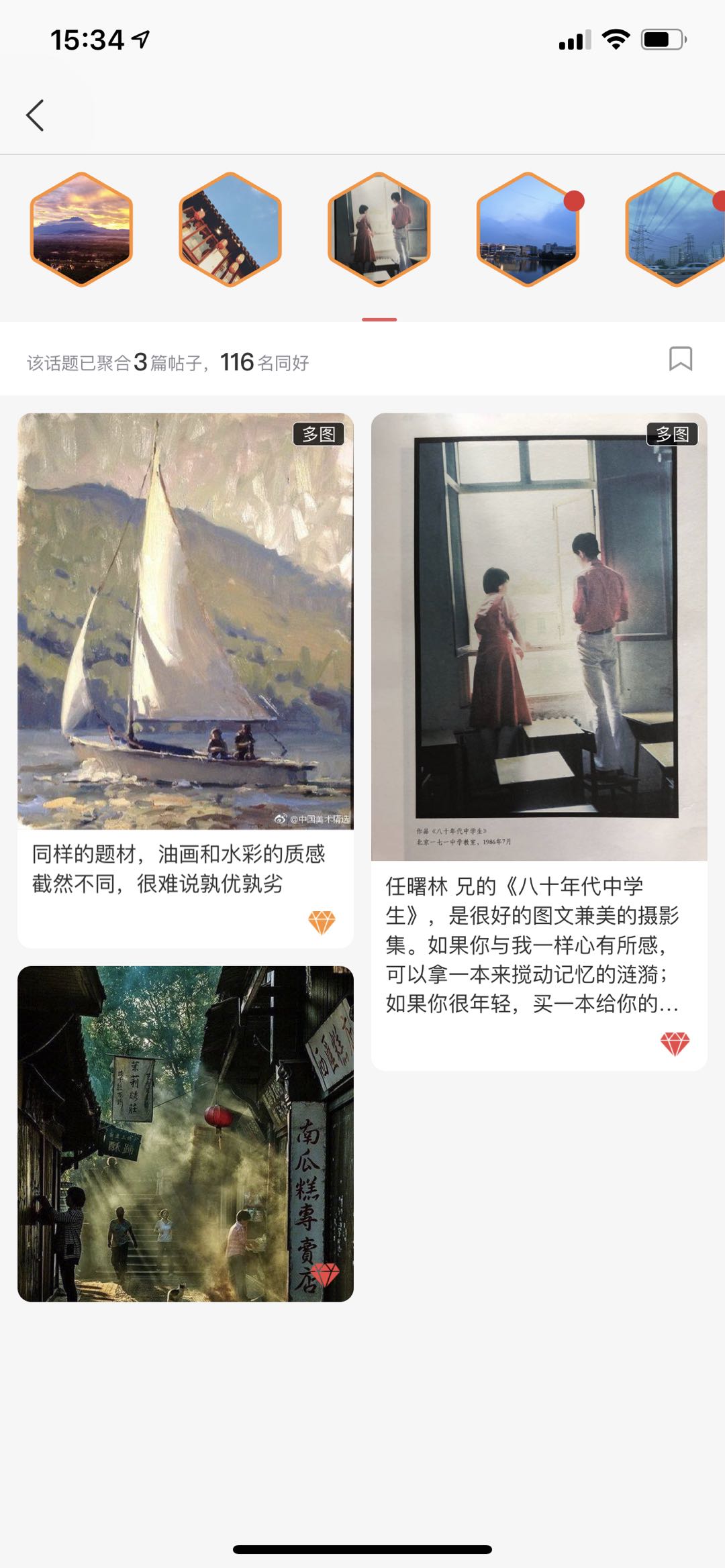}
\end{minipage}%
}%
\centering
\caption{Four typical nodes of the KNS in VISVA, comprise 3, 3, 12, and 3 files, respectively.}\label{Fig:visva_node}
\end{figure}

Fig. \ref{Fig:visva_node} shows four typical knowledge items\footnote{The results are captured from the VISVA China Ta Zai APP.}, consisting of 3, 3, 12, and 3 files, respectively. We can feel that the files in each group share some common features, but from different perspectives. For instance, the first 3 files somehow share the units "blue, water, ocean", while the third 12 files share the same author, the official account of VISVA China\footnote{Without using any tag or particular channel, it is surprising to see the system can group the 12 files together without any \emph{noise}.}. Meanwhile, we can feel sometimes, such as the first and the fourth one, that the common units in an item are hard to precisely described in word, so we may regard these knowledge items as \emph{mind} knowledges\footnote{Actually, the vision of VISVA is to construct a global brain \cite{bloom2000global}.}, which is far beyond the conventional entry system.

As for the SNR, in practice we use $$SNR = \frac{\#\text{clicked files}}{\#\text{pushed files}}$$ instead of eqn. (\ref{eqn:snr_a}) to measure it, by simply assuming that once a user clicked a file, he/she is interested in it, and vice versa. By the VISVA system, the recommendation SNR is promoted from around $0.1$, the random recommendation SNR, to around $0.35$, that is, VISVA significantly improves the recommendation qualify by 3.5 times\footnote{By comparison, in an artificial test of VISVA, by setting $|A|=1000, |U|=54$, number of files $|F|=953$, and average $\frac{1}{953}\sum_{i=1}^{953}f_i^Tf_i=3.178, \frac{1}{1000}\sum_{i=1}^{1000}a_i^Ta_i = 1.543$, where $f, a$ is defined by eqn.(\ref{eqn_fa}), the system SNR was greatly improved from the random $0.0883$ to $0.5492$.}.


\subsection{other potential applications}
The KNS and ECI can also be straightforwardly applied to some other types of recommendation system, such as the electric commercial platform to achieve more reasonable product recommendations, or it can be served as a fundamental multi-platform (commercial, reading, education, restaurant etc.) information spreading system, to integrate these looking-very-different information and realize a cross-cutting recommendation at the operation system (OS) level.

From the perspective of information flow, the KNS and ECI realize a high efficient information distribution from the sender to target receivers. So the KNS in this sense functions as a router, which is a core requirement for the internet of things (IoT). From the perspective of machine learning, the ECI realizes an automated feature extraction or data clustering (by the nodes in KNS), two important tasks in machine learning. Though there are further more challenges to be solved (see the last section for details), the KNS and ECI do provide a new perspective and tool to re-consider and re-formulate these problems.

\section{Conclusions and Future Works}\label{sec_6}
In this article we introduced the KNS and ECI algorithm. Basically, the ECI algorithm collectively takes advantages of intelligence of agents in an evolutionary manner, thus named ECI. The ECI based KNS is a new knowledge generation, storage and distribution system, which avoids many drawbacks and bottlenecks encountered by nowadays knowledge, recommendation, search and social systems. A real application, VISVA, was built based on the KNS and ECI, and has exhibited some promising and quite surprising results. On the other hand, this paper basically gives only some preliminary theoretical analysis about the model and algorithm, on which there are many open problems worth further exploring.

First, as an original but mainly heuristical algorithm, though has achieved great progresses especially in real application, the ECI still needs further theoretical analysis, on such as convergence, error bound, and complexity. In addition to some theoretical insights, it will also definitely help to further improve the algorithm itself.

Second, it is also beneficial to make an extensive comparison, from the perspective of model and algorithm instead of just function, with related works. (Google) knowledge graph and collaborative filtering \cite{herlocker1999algorithmic} used by recommendation system, the Pagerank \cite{page1999pagerank} like voting techniques used by search engine, are typical examples.

Last but not least, how to design machine agent for ECI is another important further work. By machine agents, which might be preset or learned during the evolution, the information generation and distribution mechanisms of KNS and ECI may generalize them to vast applications in the fields of, such as, machine learning, evolutionary computation and IoT.

\end{document}